\documentstyle[11pt,paspconf, epsf]{article}

\begin{document}

\title{Chemical Models of Collapsing Envelopes}
\author{Edwin A. Bergin}
\affil{Harvard-Smithsonian Center for Astrophysics, Cambridge, MA 02138}

\begin{abstract}
We discuss recent models of chemical evolution in the developing and
collapsing protostellar envelopes associated with low-mass star formation.
In particular, the effects of depletion of gas-phase molecules onto grain surfaces
is considered.  
We show that during the middle to late
evolutionary stages, prior to the formation of a protostar, various
species selectively deplete from the gas phase.   The principal pattern of
selective depletions is the depletion of sulfur-bearing molecules relative to 
nitrogen-bearing species: NH$_3$ and N$_2$H$^{+}$.  This pattern 
is shown to be insensitive to the details of the dynamics 
and marginally sensitive to whether the grain mantle is dominated
by polar or non-polar molecules.
Based on these results we 
suggest that molecular ions are good tracers of collapsing
envelopes.
The effects of coupling chemistry and
dynamics on the resulting physical evolution are also examined. 
Particular attention is paid to comparisons between
models and observations.   

\end{abstract}


\section{Introduction}

The formation of dense molecular condensations, and of stars, 
involves large changes in the physical properties of initially atomic
and eventually molecular gas.   These changes have very specific
consequences on the chemical interactions of the gas and dust inside the forming
cores/stars.   In particular, as the density increases molecules in the
gas phase will collide with dust grains with greater frequency and, if the
molecules stick with any reasonable efficiency, they will deplete from the  
gas phase.  Indeed, molecular depletions have been suggested as the 
primary chemical indicator of the star formation process 
(c.f. Mundy \& McMullin 1997).  

Chemical models of the collapsing envelopes associated with star formation 
have begun to examine in detail how this process alters the chemistry.  These
models have focused initially on how to maintain the gas-phase chemistry observed
in dense cores, but also on the specific patterns of molecular
depletions.   This pattern could simply consist of all species depleting on grains
at their respective depletion timescales or, with some desorption,
removal from the grain surfaces could be selective in nature with some molecules disappearing from the
gas phase at early stages while others deplete at later stages.
If such a pattern could be isolated, then observations of chemical structure
could potentially be used as an
indicator of the various phases of the star formation process.

In this paper we discuss recent models of the ``chemistry of star formation''.  We predominantly
focus on chemical models which examine the earliest phases of low mass star formation: collapse of
low density molecular envelopes into isolated dense cores and single stars.    For a detailed
discussion of chemical models for high (and low) mass star formation the reader is referred to the 
recent review by van Dishoeck and Blake (1998).

\section{Observations of Molecular Depletions}

The best method to 
search for molecular depletions is to directly observe molecules in the solid
state.  This can be done from the ground in many instances (Whittet 1993), but the
{\em Infrared Space Observatory} has provided the clearest picture to date of the dominant
molecules on molecular grain mantles (see Ehrenfreund, this volume).  However, this method
fails if  wish to investigate what happens to the observed gas-phase in molecular clouds as a
core condenses.  The molecular emission that we observe in star forming regions is from molecules
such as CS and NH$_3$, which have abundances as low as 10$^{-8}$ relative to H$_2$.
Such low abundances, if completely frozen in grain mantles, would produce 
absorption features 
to weak to observe.  Thus, we are reduced to indirect methods of observing
molecules in the gas-phase and searching for emission changes, which could potentially be
attributed to depletion.  Here, we suffer the additional
difficulty of the similar effects on molecular excitation due to changes in abundance or  
in the physical conditions. 

Some careful excitation analyses have been performed and it is certain the
molecular depletions have been isolated in several objects.  The most striking example is in the
starless core L1498
where Kuiper et al (1996) suggest that CS and C$_2$S are either diminished in abundance or
completely absent in the dense central regions, while observations of
NH$_3$ show centrally condensed emission.   A similar case is in L1544, also a starless
core, where
C$_2$S emission appears to anti-correlate with both dust continuum emission and N$_2$H$^{+}$ emission
(Ohashi et al 1999; Ohashi 2000, this volume).
An excitation analysis using N$_2$H$^{+}$ shows that this effect is not due to excitation
and is likely the result of C$_2$S depletion in the dense core traced by dust and N$_2$H$^{+}$ (Plume et al 2000).
Depletion of CO has also been isolated in a cold dense core in IC5146 (Kramer et al 1999), while
similar trends are found in more evolved sources
(Blake, van Dishoeck, \& Sargent 1992).
For a review of molecular depletions from
an observational perspective see Mundy \& McMullin (1997) or Hogerheijde (this volume).

\section{Dynamical Models}

To model the chemical interactions associated with low mass star formation we should look 
to the current paradigm for the dynamical evolution of collapse or the so called ``standard model
of star formation'':   

\begin{itemize}
\begin{enumerate}
\item A magnetically supported core condenses quasistatically via ambipolar diffusion 
(Basu \& Mouschovias 1994; Lizano \& Shu 1989).

\item During collapse the outer layers approach a $\rho \propto r^{-2}$ density distribution,
which is simply hydrostatic equilibrium.

\item As the density increases in the center, the ionization fraction decreases,
and the field lines decouple from the gas.

\item Collapse then ensues in an inside-out fashion (Shu 1977).
\end{enumerate}
\end{itemize}

\noindent An excellent review of our present understanding of star formation can be
found in Hartmann (1998).  For critical discussions of the standard model see
Whitworth et al (1994) and Nakano (1998).   

\section{Gas-Grain Interactions}

An important part of any coupled chemical and dynamical models 
is physics of the interaction between molecules and grain surfaces.   
Two of the most relevant processes are the sticking coefficient and the binding
energy.
(1) The sticking coefficient is simply how often a molecule will stick to the
surface after a collision with a grain and is 
typically assumed to be between 0.1 and 1. The theory behind this 
assumption is discussed by Williams (1993).
(2) The process by which molecules are adsorbed or bound onto grain surfaces is known as physical adsorption.
When a molecule approaches a grain there is an induced dipole-dipole (van der Waals) interaction between the 
gas phase molecule and whatever species is dominating the grain surface, either an exposed
silicate surface,
or more likely H$_2$O or CO molecules.
This interaction results in an attractive
force which binds the molecule to the surface in a potential well
(e.g. Tielens \& Allamandola 1987).  
Because this
is an induced dipole-dipole interaction the strength of the attraction depends on the polarizability
of the molecules involved, and a polar molecule, such as H$_2$O, 
presents a stronger binding surface than a surface covered by a non-polar molecule, such as CO.

\section{Chemical Models}

The large scale of the star formation problem, which requires coupling very
complex hydrodynamical codes with complicated chemical models, renders a complete
examination difficult.  The models relevant for this discussion are those that have extracted
parameterized fits to the density evolution from the dynamical simulations and used these
in concert with complex chemical codes.   Of these models, each incorporates 
gas-grain interactions, but they
vary greatly in the desorption mechanism theorized to remove molecules from the grain surface.
The models are ordered below in terms of increasing desorption efficiency.

Rawlings et al. (1992) examine the inside-out collapse solution of Shu (1977) with 
no desorption.  Bergin \& Langer (1997) use the ambipolar diffusion density
evolution given in Basu \& Mouschovias (1994), along with another dynamical solution.
These models include cosmic-ray spot heating wherein a cosmic ray strikes a grain and
evaporates off molecules from a localized hot spot near the impact.  Willacy, Rawlings, \&
Williams (1994) use the Shu solution and incorporate both cosmic-ray heating and heating
due to H$_2$ formation.  Shalabiea \& Greenberg (1997) also use the
Shu solution with efficient desorption via grain explosions.  One common assumption is that at the
start of the evolution all of the hydrogen is in molecular form.

The above models each assume isothermal evolution at 10 K and, hence, are appropriate for
the initial stages of collapse or for the outer envelopes once a protostar has formed
in the innermost regions.  Ceccarelli et al (1996) include the effects of central heating
due to the protostar and these results are discussed in \S 6.  Below we will briefly discuss
the models of Rawlings et al (1992) and, in greater detail, those of
Bergin \& Langer (1997).   

\subsection{Rawlings et al 1992}

\begin{figure}
\plotfiddle{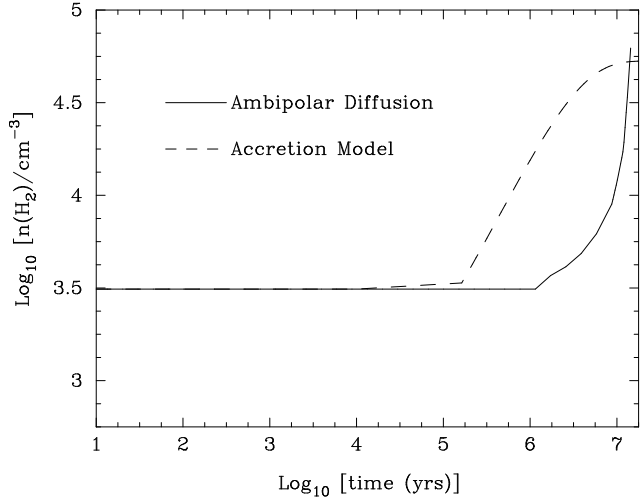}{2.4in}{0}{52}{52}{-170}{-185}
\caption{The evolution of the density of molecular hydrogen as a function  
of time for the evolution through ambipolar diffusion of Basu \& Mouschovias (1994)
(solid line) and an accretion model (dashed line).  
}
\end{figure}
These models, using the Shu (1977) inside-out collapse solution, examine the evolution at
every point in the collapsing envelope at all times, but include depletion with no desorption.
Therefore, all molecules deplete from the gas phase
as the density increases.   One interesting result of these simulations is that when the neutral
species deplete the abundances of molecular ions increases.  This is 
the result of a constant ion formation rate from cosmic rays, coupled with a decreasing 
ion destruction
rate from reactions with the depleted neutral species.  Eventually,
the molecular ions will disappear from the gas phase 
as their parent molecules deplete
(e.g. CO for HCO$^{+}$  or N$_2$ for N$_2$H$^{+}$). 
This suggests that molecular ions should
be good tracers of collapsing envelopes.

\subsection{Bergin \& Langer 1997}

This work investigates the evolution a single parcel of gas deep inside a core using
two dynamical solutions.  The first is based on the work of Basu \& Mouschovias (1994) who examined
the formation of a dense cloud core via diffusion of magnetic flux.   The second is a phenomenological
model wherein a static dense core is accreting material from a low density halo and the 
gas is brought from low to high density within a few free-fall times.  
The time evolution of density for these models is shown in Figure~1. 
The chemical models included cosmic-ray spot heating, and examined
solutions where grains are coated with a layer of water molecules (strongly binding) and, alternately,  a
layer of CO molecules (weakly binding).

For the chemical models at t = 0 carbon and metal atoms are in ionized form,
while the oxygen and nitrogen atoms are neutral.  All
species are depleted from solar abundances, with the heavy metals 
more severely depleted.
A dense pre-protostellar core forms from moderately dense gas and dust
composed of a mix of atomic and molecular species.  To establish this state the 
chemistry is allowed to evolve at constant density for 10$^5$ yrs and these abundances 
are used as initial conditions for the time-dependent density model.  

\begin{figure}
\plotfiddle{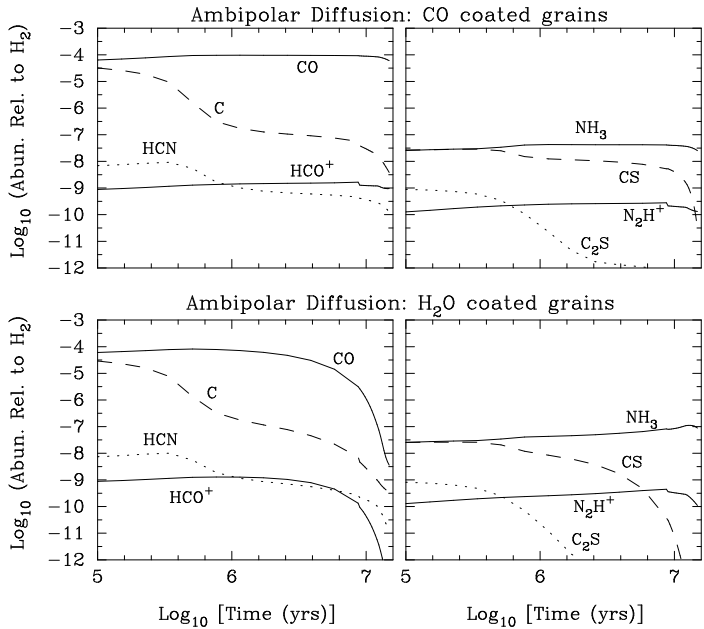}{3.0in}{0}{50}{50}{-170}{-100}
\caption{Chemical abundances relative to H$_2$
as a function of time for the ambipolar diffusion model. 
The top panel shows the chemical evolution for a model with CO-coated grains 
and the bottom for H$_2$O -coated grains.
}
\end{figure}
Chemical abundances as a function of time for the first model of dynamical evolution
via diffusion of magnetic flux are shown in Figure~2.  
In the model with CO-coated grains, 
abundances exhibit only small changes until t $\sim 3 \times 10^{5}$ yrs.
At this time the abundances of simple and complex carbon-bearing species begin to decline rapidly.  
This change is not the
result of any density increase 
(see Figure~1).  Rather the rapid decline in the abundances of 
HCN and C$_2$S is related to the sharp drop in the abundance of neutral carbon resulting from the
onset of CO formation.  There is also no discernible effect of molecular depletion until
t $\sim 10^{7}$ yr (n$_{H_2}$ $> 10^4$ cm$^{-3}$) when 
the abundances of CS and C$_2$S drop precipitously as these molecules deplete onto
grain surfaces.  Even at late evolutionary stages a reasonable amount of CO, HCO$^{+}$, HCN,
NH$_3$, and N$_2$H$^{+}$ remain in the gas.  

This depletion pattern is reminiscent of the chemical structure 
observed in L1498, in L1544, and in other cores (see \S3; Ohashi, this volume) and is simply
understood by the way the various atomic pools interact with the grains.   For the 
C, O, and N pools the dominant species are highly non-polar (CO, O, O$_2$, N, N$_2$) and,
while these species are depleting onto grains, cosmic-ray desorption is able to
remove them from the weakly binding CO surface.  These molecules are processed in the gas phase
by cosmic-ray induced photons and reactions with He$^{+}$ atoms, eventually creating 
complex polar molecules that will deplete such as HCN or H$_2$CO.  
However, the dominant species (e.g. CO, N$_2$) are more abundant (by three or four orders of magnitude) 
than more complex molecules
and essentially carry the chemistry on their backs.  The opposite is true for the sulfur chemistry, 
the dominant molecules are CS and SO, which are polar molecules with high binding potentials 
(even on a CO surface).  These species deplete onto grain
surfaces and are very difficult to remove.  
Hence, the sulfur chemistry is halted while the
C, O, and N chemistry continues until densities reach levels that depletion
overwhelms desorption.

Figure~2 also shows the evolution of the ambipolar diffusion model for the case where the 
grain mantles are dominated by water ice.  Here the binding energies are much
higher than in the previous case and most molecules disappear from the 
gas phase, except for  N$_2$H$^{+}$ and NH$_3$.
This result is interesting and is due to the fact that 
N$_2$, the parent molecule of both N$_2$H$^{+}$ and NH$_3$,
is one of the most volatile molecules in star forming clouds. 
Thus, N$_2$H$^{+}$ and NH$_3$ are suggested by these models to be the best tracers of dense collapsing 
gas. 

The above models examined the chemistry using the dynamical solutions appropriate for
core formation via the diffusion of magnetic flux.  In Figure~3 we examine the effect of 
changing the dynamical solution.  In the figure we show the 
previous model of ambipolar diffusion side-by-side with the accretion model discussed earlier.  We only
present results for the model with CO-coated grains, but similar effects are found 
for water ice dominated mantles (see Bergin \& Langer 1997).  We find
little change in the carbon chemistry between the solutions (top panels).  For the
sulfur and nitrogen chemistry, presented in the bottom panels, there are some differences, 
mainly that CS begins to
deplete from the gas phase at earlier times in the accretion model.  The explanation for this
difference can be found in Figure~1 where we see that the gas spends a larger amount of time
in a high density state in the accretion model than for ambipolar diffusion, which results in a
shorter depletion timescale for CS.  

In general,
we conclude that changing the dynamical model has little effect on the solution.
This result is quite robust in the sense that similar results would be found
for any dynamical model that increases with time or is centrally concentrated.
Therefore, the density gradient that  results from core condensation
is accompanied by chemical gradients, with the inner parts of the
core representing high density chemistry (with significant CS and CCS depletions onto grains and
N$_2$H$^{+}$ remaining in the gas phase)
and the outer parts representing the original low density molecular composition.
The size scale of the CS or CCS ``hole'' will primarily depend on the radius where
the gas density is $^{>}_{\sim}$$2 \times 10^4$ cm$^{-3}$, which is roughly the density
at which the sulfur molecules deplete. 
Using the density profiles of B335
from Zhou et al (1993) or L1689B from  Andr\'e, Ward-Thompson, \& Motte (1996),
we estimate that sulfur depletion occurs at radii $< 0.025$ pc (or $\sim$30--40$''$ at 
the distance of Taurus).  This is comparable to the size of the CCS depletion zone
observed in L1498 and L1544 (Kuiper et al 1996; Ohashi et al 1999).
More efficient desorption might change these results, but observations
of molecular depletions are beginning to rule out a very efficient mechanism.
However, these results would suggest that chemistry cannot be used to 
help constrain the dynamics and that the assignment of a true ``chemical'' 
timescale awaits a better understanding of the dynamics.

\begin{figure}
\plotfiddle{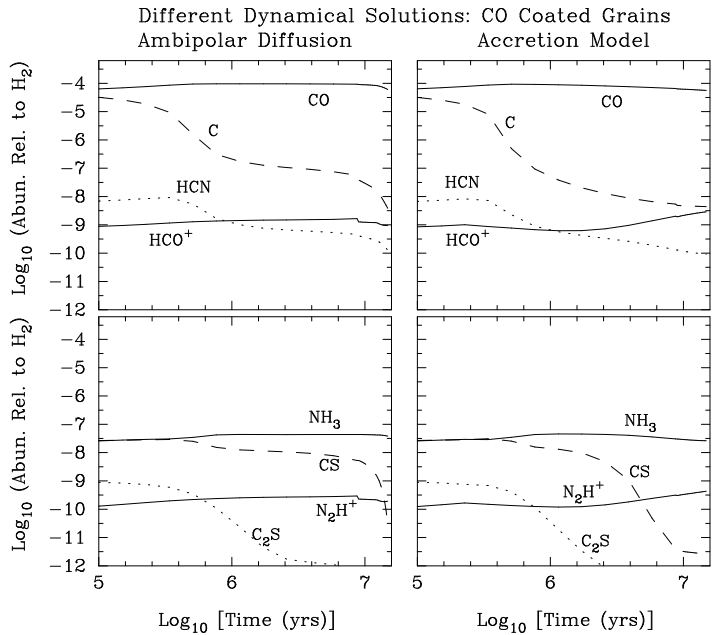}{3.0in}{0}{50}{50}{-170}{-105}
\caption{The chemical abundances relative to H$_2$
as a function of time comparing the two dynamical solutions: ambipolar diffusion model (left-hand panels),
accretion model (right-hand panels).   These models are for CO-coated grains.
}
\end{figure}

\section{Chemistry and Star Formation}

\begin{figure}
\plotfiddle{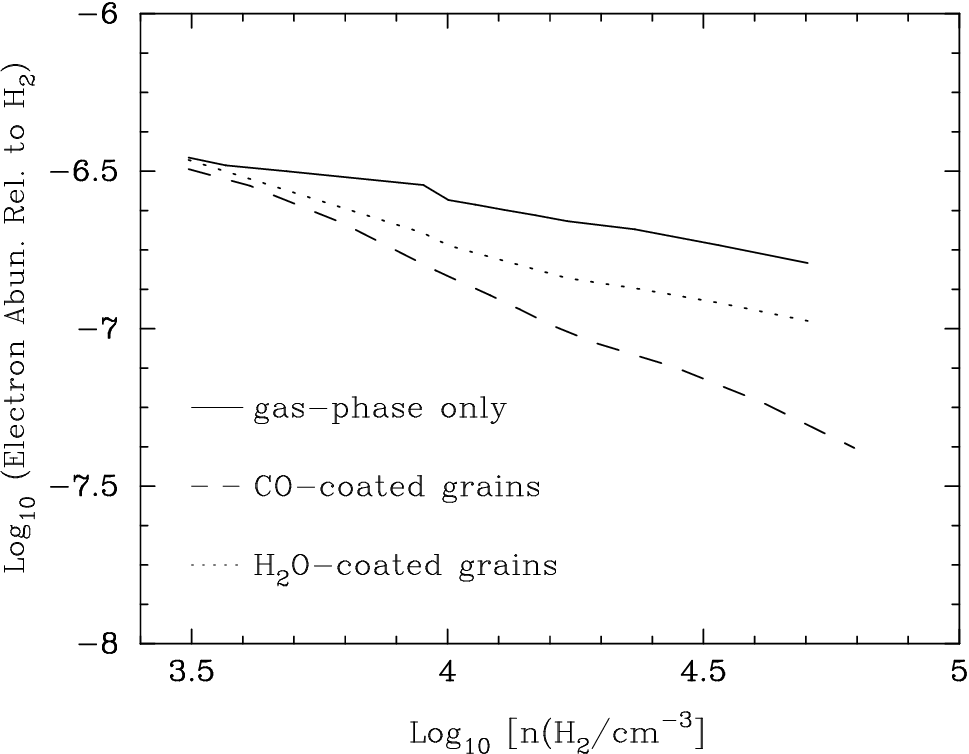}{2.4in}{0}{52}{52}{-170}{-185}
\caption{The chemical abundances relative to H$_2$
as a function of time comparing the two dynamical solutions: ambipolar diffusion model (left-hand panels),
accretion model (right-hand panels).   These models are for CO-coated grains.
}
\end{figure}

\begin{table}
\begin{center}\footnotesize
\begin{tabular}{llll}
\multicolumn{4}{c}{TABLE 1} \\ 
\multicolumn{4}{c}{Chemical Evolution and Star Formation} \\ \\
\hline
\multicolumn{1}{c}{Evolutionary Phase} &
\multicolumn{1}{c}{Molecular Abundances} &
\multicolumn{1}{c}{n$_{H_{2}}$ (cm$^{-3}$)} &
\multicolumn{1}{c}{Example} \\\hline 
Dense Core Formation & Normal Abundances & $\sim$3 $\times 10^{3}$& HCL2 \\\\
Evolving Cores: &  &  & \\\
(Early Stages) & Little CS and C$_2$S depletion & $\sim 10^{4}$ & TMC-1 \\\
(Late Stages) & Increasing CS and C$_2$S depletion & $\sim$ 3 $\times 10^{4}$ & L1498 \\
& N$_2$H$^{+}$ and HCO$^{+}$ tracing envelope & & \\ \\
Collapsing Phase & Significant CS and C$_2$S depletion & $>$ 10$^{5}$ & (?) \\
& HCO$^{+}$ depleting prior to N$_2$H$^{+}$ &  & \\ \\
Protostar Formation & & & \\
(outer envelope) & Significant depletion, strongly & $> 10^{6}$ & HL Tau, \\
& dependent on mantle properties & & DG Tau\\ 
(inner envelope) & release of mantle species, & $> 10^{7}$ & IRAS16293 \\
stellar heating & high-T chemistry (H$_2$O), & & \\
& depression of  HCO$^{+}$ and N$_2$H$^{+}$ && \\
& abundances at center && \\ \\
Onset of Wind Activity & Full liberation of grain mantles & & Orion IrC2 \\
& near star, shocked regions (H$_2$O and SiO)  && \\\\
Late Embedded Period & Normal Abundances && \\\hline
\end{tabular}
\end{center}
\end{table}

It appears clear from the preceding discussion that the process of star formation
does have significant and potentially observable effects on the chemistry of the gas.
However, chemistry is also an important part of the dynamical evolution as well.
As an example in Figure~4  we show the ionization fraction as a function
of density taken from Bergin \& Langer (1997),
using the ambipolar diffusion dynamical solutions of Basu \& Mouchovias (1994).
A pure gas-phase model is shown, in addition to gas-grain simulations using CO and H$_2$O mantles,
Clearly the electron abundance is a function of density, however, what is new in these results
is that the character of the density dependence changes with the strength of the 
gas-grain interaction.   In this instance, with the density evolution
taken from a dynamical model driven by ion-neutral drift, such differences could be
important, which is an excellent argument for the incorporation of chemical interactions into 
dynamical models.

In summary, over the past several years the wealth of observational data  and 
theoretical chemical simulations has provided our field with important clues 
to the puzzle of star formation.  At the previous IAU conference Mundy \& McMullin (1997)
provided a general outline of the chemical evolutionary sequence associated with each
phase of the star formation process.  We believe that the observations and theory are
converging to a point that allows for greater specification.  We provide a more
detailed evolutionary sequence in
Table~1.  The dynamical evolution begins with a core at low density with
normal abundances.  As the core evolves the principal chemical signature of condensation
is the depletion
of sulfur-bearing molecules relative to the nitrogen-bearing species, NH$_3$ and
N$_2$H$^+$.  Thus, for early stages, prior to collapse CS or C$_2$S are good probes;
during collapse the best tracers of the cold dense envelope are the molecular ions:  N$_2$H$^+$ and
HCO$^{+}$.  At later evolutionary stages HCO$^+$ should deplete prior to N$_2$H$^+$,
making N$_2$H$^+$ the most robust tracer of dense gas.  
The above conclusion is not applicable for all stages, once a star forms in the center
it will locally heat the surrounding material releasing mantle species and, perhaps,
allowing for high temperature chemistry to create water (Ceccarelli et al 1996).  The
flood of neutrals into the gas phase will destroy molecular ions, making neutral molecules
the better tracers for the inner envelope.   At the final stages, developing 
winds will interact with surrounding gas producing
shocked regions which can be probed by SiO and H$_2$O.


\acknowledgments

The author is immensely grateful for the help, advice, and fruitful collaboration with W. Langer,
and to D. Wilner for a careful reading of the manuscript.


\begin{references}
\reference Basu, S. \& Mouschovias, T. Ch. 1994, \apj, 432, 720 
\reference Bergin, E.A. \& Langer, W.D. 1997, \apj, 486, 316
\reference Blake, G.A., van Dishoeck, E.F., \& Sargent, A. I. 1992, ApJ, 391, L99
\reference Ceccarelli, C., Hollenbach, D.J., \& Tielens, A.G.G.M. 1996, \apj, 471, 400
\reference Hartmann, L. 1998, Accretion Processes in Star Formation, Cambridge Univ. Press: Cambridge
\reference  Kramer, C. et al. 1998b, A\&A, 342, 257
\reference Kuiper, T. B. H., Langer, W. D. \& Velusamy, T. 1996, \apj, 468, 761
\reference Lizano, S., \& Shu, F. II 1989, \apj, 342, 834
\reference Mundy, L.G., \& McMullin, J.P. 1997, in Molecules in Astrophysics: Probes and Processes,
E.F. van Dishoeck, Kluwer: Dordrecht, 183
\reference Nakano, T. 1998, \apj, 494, 587
\reference Ohashi, N., Lee, S.W., Wilner, D.J., \& Hayashi, M. 1999, \apj, 518, L41
\reference Plume, R., Bergin E.A, Caselli, P., \& Myers, P.C. 2000, in prepartion
\reference Rawlings, J. M. C., Hartquist, Menten, K. M. \& Williams, D. A. 1992, \mnras, 255, 471
\reference Shalabiea, O.M., \& Greenberg, J.M. 1995, \aap, 303, 233
\reference Shu, F. II 1977, \apj, 214, 488
\reference Tielens, A. G. G. M., \& Allamandola, L. J., in Interstellar
Processes, D. J. Hollenbach \& H. A. Thronson, Reidel:Dordrecht, 397
\reference Whittet, D. C. B. 1993, in Dust and Chemistry in Astronomy,
T. J. Millar \& D. A. Williams,  Institute of Physics Publishing:Bristol, 9
\reference Whitworth, A.P., Bhattal, A.S., Francis, N., \& Watkins, S.J. 1996, \mnras, 283, 1061
\reference Willacy, K., Rawlings, J.M.C., \& Williams, D.A. 1994, \mnras, 269, 921
\reference Williams, D. A. 1993,
in Dust and Chemistry in Astronomy,
T. J. Millar \& D. A. Williams, Institute of Physics Publishing:Bristol, 143
\reference van Dishoeck, E.F., \& Blake, G.A. 1998, \araa, 36, 317
\reference  Zhou, S., Evans, N. J., Koempe, C., \& Walmsley, C. M. 1993,
ApJ, 404, 232



\end{references}
\end{document}